\documentclass[a4paper,aps,preprintnumbers,showpacs,amsmath,amssymb,prl]{revtex4}
\usepackage{mathrsfs}
\usepackage{amsmath}
\usepackage[dvips]{graphicx}
\usepackage{epsfig}
\usepackage[dvips]{graphics,graphicx}
\begin{document}

\title{Entangling two Bose Einstein condensates in a double cavity system}

\author{Tarun Kumar$^{1}$,Aranya B.\ Bhattacherjee$^{2}$ and ManMohan$^{1}$}

\address{$^{1}$Department of Physics and Astrophysics, University of Delhi, Delhi-110007, India} \address{$^{2}$Department of Physics, ARSD College, University of Delhi (South Campus), New Delhi-110021, India}

\begin{abstract}
We propose a scheme to transfer the quantum state of light fields to the collective density excitations of a Bose Einstein condensate (BEC) in a cavity. This scheme allows to entangle two BECs in a double cavity setup by transferring the quantum entanglement of two light fields produced from a nondegenerate parametric amplifier (NOPA) to the collective density excitations of the two BECs. An EPR state of the collective density excitations can be created by a judicious choice of the system parameters.
\end{abstract}


\maketitle

\section{Introduction}

 It is widely accepted that quantum entanglement is an essential ingredient for the implementation of quantum information processing devices \cite{werner}. Creating highly entangled multi-particle states is therefore one of the most challenging goals of modern experimental quantum mechanics. Quantum entanglement of two or more systems leads to correlations between observables of the systems that cannot be explained on the basis of local realistic theories. Quantum entanglement lies at the heart of the difference between the quantum and the classical world.  Due to its vast application, quantum entanglement has been studied in different systems such as optomechanical systems\cite{mancini}, Bose Einstein condensates (BEC) trapped in double quantum well \cite{chen}and in optical lattices \cite{you}. Entanglement has also been observed experimentally at NIST group \cite{sackett}, where entanglement state of four-ions was created successfully by using the scheme proposed by Sorensen and Molmer \cite{sorensen}.

 In order to build a quantum information network using atomic systems, quantum state exchange between light and matter is an essential ingredient. One prefers matter over photons as quantum memory elements since experimentally it is difficult to localize and store photons. Stimulated Raman adiabatic passage is considered to be a very convenient technique of storing light and has been used in single atom cavity quantum electrodynamics for transfer of qubits between atoms and photons and for building quantum logic gates \cite{parkins}. Mapping quantum states to collective atomic spin systems \cite{kozhekin} and quantized vibrational states of trapped atoms \cite{kimble} by means of stimulated Raman adiabatic passage technology and electromagnetically induced transparency \cite{lukin} has been discussed earlier.

 New possibilities for cavity opto-mechanics may emerge by combining the tools of cavity quantum electrodynamics(QED) with those of ultracold gases \cite{brennecke, murch, bhattacherjee09}. Placing an ensemble of atoms inside a high-finesse cavity enhances the atom-light interaction because the atoms collectively couple to the same light mode. The motional degrees of freedom of ultracold atomic gases represent a new source of long-lived coherence affecting light-atom interaction. Nonlinear optics arising from this long-lived coherent motion of ultracold atoms trapped within a high-finesse Fabry-Perot cavity was reported recently \cite{murch}. Strong optical nonlinearities were observed even at low mean photon number of $0.05$. This nonlinearity also gives rise to bistability in the transmitted probe light through the cavity.
 Experimental implementation of a combination of cold atoms and cavity QED (quantum electrodynamics) has made significant progress \cite{Nagorny03,Sauer04,Anton05}. Theoretically there have  been some interesting work on the correlated atom-field dynamics in a cavity. It has been shown that the strong coupling of the condensed atoms to the cavity mode changes the resonance frequency of the cavity \cite{Horak00}. Finite cavity response times lead to damping of the coupled atom-field excitations \cite{Horak01}. The driving field in the cavity can significantly enhance the localization and the cooling properties of the system\cite{Griessner04,Maschler04}. It has been shown that in a cavity the atomic back action on the field introduces atom-field entanglement which modifies the associated quantum phase transition \cite{Maschler05}. The light field and the atoms become strongly entangled if the latter are in a superfluid state, in which case the photon statistics typically exhibits complicated multimodal structures \cite{Chen07}. A coherent control over the superfluid properties of the BEC can also be achieved with the cavity and pump \cite{Bhattacherjee}. Recently, a new approach was proposed which is based on all optical measurements that conserve the number of atoms. It was shown that atomic quantum statistics can be mapped on transmission spectra of high-Q cavities, where atoms create a quantum refractive index. This was shown to be useful for studying phase transitions between Mott insulator and superfluid states since various phases show qualitatively distinct spectra \cite{Mekhov07}. Motivated by such interesting developments in the field of cavity electrodynamics of Bose Einstein condensates, we propose in this study a new scheme of entangling two Bose-Einstein condensates (BEC) in a double cavity set-up by transferring the entanglement of quantum-correlated light fields produced from a nondegenerate parametric amplifier (NOPA) to the collective density excitations of the two BECs.

\section{Quantum state transfer from light to BEC}

 In this section, we show that the quantum state of the input light field can be transferred to a BEC trapped in a cavity. We consider an elongated cigar shaped Bose-Einstein condensate(BEC) of $N$ two-level $^{87} Rb$ atoms in the $|F=1>$ state with mass $m$ and frequency $\omega_{a}$ of the $|F=1>\rightarrow |F'=2>$ transition of the $D_{2}$ line of $^{87} Rb$, strongly interacting with a quantized single standing wave cavity mode of frequency $\omega_{c}$ as shown in Fig. 1. The standing wave that forms in the cavity results in an one-dimensional optical lattice potential. The cavity field is also coupled to external fields incident from one side mirror.  It is well known that high-Q optical cavities can significantly isolate the system from its environment, thus strongly reducing decoherence and ensuring that the light field remains quantum-mechanical for the duration of the experiment. We also assume that the induced resonance frequency shift of the cavity is much smaller than the longitudinal mode spacing, so that we restrict the model to a single longitudinal mode.  In order to create an elongated BEC, the frequency of the harmonic trap along the transverse direction should be much larger than one in the axial (along the direction of the optical lattice) direction. The system is also coherently driven by a laser field with frequency $\omega_{p}$ through the cavity mirror. Here the two mirrors $M_{1}$ and $M_{2}$ are fixed. This system is modelled by a Jaynes-Cummings type Hamiltonian ($H_{JC}$) in the rotating and dipole approximation.

 \begin{figure}[t]
\hspace{-0.0cm}
\includegraphics [scale=0.5]{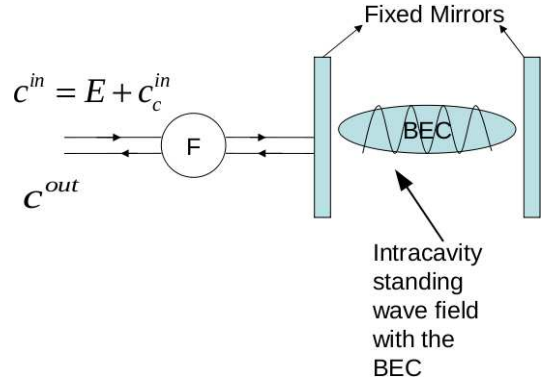}
\caption{Schematic diagram of the BEC in a cavity coupled with the cavity field. Both the mirrors are fixed. A pump light $c^{in}$ is incident from the left mirror and from the same mirror the reflected light from the BEC is taken out as $c^{out}$. Faraday oscillator ($F$) enables a unidirectional coupling. }
\label{1}
\end{figure}

\begin{equation}
H_{JC}=\frac{P^{2}}{2m}-\hbar \Delta_{a} \sigma^{+} \sigma^{-}+\hbar \Delta_{c}c^{\dagger} c-i \hbar g(x)[\sigma^{+}c-\sigma^{-}c^{+}]+ i \hbar \sqrt{2 \gamma}(c_{in}c^{\dagger}-c_{in }^{*} c ),
\end{equation}

where $\sigma^{+}$ and $\sigma^{-}$ are the Pauli matrices and $\Delta_{a}=\omega_{p}-\omega_{a}$ and $\Delta_{c}=\omega_{c}-\omega_{p}$ are the large atom-pump and cavity-pump detuning, respectively. . $g(x)=g_{o} \cos{kx}$ is the space dependent atom field coupling. $c$ and $c^{\dagger}$ are the annihilation and creation operators of the intracavity field. The input laser field populates the intracavity mode which couples to the atoms through the dipole interaction. The field in turn is modified by the back-action of the atoms. The system is open as the cavity field is damped by the photon leakage through the coupling mirror. We are considering a system with large detuning and hence spontaneous emission is ignored and hence we can adiabatically eliminate the excited state using Heisenberg equation of motion $\dot \sigma^{-}=\frac{i}{\hbar}[H_{JC}, \sigma^{-}]$, which yields the single particle Hamiltonian

\begin{equation}
H_{o}= \frac{P^{2}}{2m}+\hbar \Delta_{c} c^{\dagger} c+ i \hbar \sqrt{2 \gamma}(c_{in} c^{\dagger}-c_{in }^{*} c)+\cos^{2}{kx}[V_{cl}+\hbar U_{o} c^{\dagger} c]
\end{equation}

The parameter $U_{0}=\dfrac{g_{0}^{2}}{\Delta_{a}}$ is the optical lattice barrier height per photon and represents the atomic backaction on the field . Also $V_{\text{cl}}({\bf{r}})$ is the external classical potential. Here we will always take $U_{0}>0$. In this case the condensate is attracted to the nodes of the light field and hence the lowest bound state is localized at these positions which leads to a reduced coupling of the condensate to the cavity compared to that for $U_{0}<0$.  Along $x$, the cavity field forms an optical lattice potential of period $\lambda/2$ and depth ($\hbar U_{0}<\hat{a}^{\dagger}\hat{a}>+V_{cl}$). We now write the Hamiltonian in a second quantized form including the two body interaction term.

\begin{eqnarray}
H&=&\int d^3 x \Psi^{\dagger}(\vec{r})H_{0}\Psi(\vec{r})\nonumber \\&+&\dfrac{1}{2}\dfrac{4\pi a_{s}\hbar^{2}}{m}\int d^3 x \Psi^{\dagger}(\vec{r})\Psi^{\dagger}(\vec{r})\Psi(\vec{r})\Psi(\vec{r})\;
\end{eqnarray}

where $\Psi(\vec{r})$ is the field operator for the atoms. Here $a_{s}$ is the two body $s$-wave scattering length. The corresponding opto-mechanical-Bose-Hubbard (OMBH) Hamiltonian can be derived by writing $\Psi(\vec{r})=\sum_{j} \hat{b}_{j} w(\vec{r}-\vec{r}_{j})$, where $w(\vec{r}-\vec{r}_{j})$ is the Wannier function and $\hat{b}_{j}$ is the corresponding annihilation operator for the bosonic atom at the $j^{th}$ site. Retaining only the lowest band with nearest neighbor interaction, we have

\begin{equation}
H= E_{o} \sum_{i} b_{i}^{\dagger} b_{i} +\hbar \Delta_{c} c^{\dagger} c+ i \hbar \sqrt{2 \gamma}(c_{in} c^{\dagger}-c_{in }^{*} c)+[V_{cl}+\hbar U_{o} c^{\dagger} c] J_{o} \sum_{i}  b_{i}^{\dagger} b_{i}+\frac{U}{2} \sum_{i} b_{i}^{\dagger}b_{i}^{\dagger} b_{i} b_{i}
\end{equation}

where

\begin{eqnarray}
U&=&\dfrac{4\pi a_{s}\hbar^{2}}{m}\int d^3 x|w(\vec{r})|^{4}\nonumber \\
E_{0}&=&\int d^3 x w(\vec{r}-\vec{r}_{j})\left\lbrace \left( -\dfrac{\hbar^2 \nabla^{2}}{2m}\right) \right\rbrace w(\vec{r}-\vec{r}_{j})\nonumber \\
J_{0}&=&\int d^3 x w(\vec{r}-\vec{r}_{j}) \cos^2(kx)w(\vec{r}-\vec{r}_{j}).
\end{eqnarray}

Here, we have neglected the tunneling terms $E=\int d^3 x w(\vec{r}-\vec{r}_{j})\left\lbrace \left( -\dfrac{\hbar^2 \nabla^{2}}{2m}\right) \right\rbrace w(\vec{r}-\vec{r}_{j \pm 1})$ and
$J=\int d^3 x w(\vec{r}-\vec{r}_{j}) \cos^2(kx)w(\vec{r}-\vec{r}_{j \pm 1})$, as they can be made small by varying the optical lattice depth. Under this approximation, the matter-wave dynamics is not essential for light scattering. In experiments, such a situation can be realized because the time scale of light measurements can be much faster than the time scale of atomic tunneling.

The dynamics of the system can be described by the following quantum Langevin equations

\begin{equation}
\dot c=-i \Delta_{c} c-i U_{o}c J_{o} \sum_{i} b^{\dagger}_{i} b_{i}+\sqrt{2 \gamma}c^{in}-\gamma c
\end{equation}

\begin{equation}
\dot b_{i}=-\frac{i E_{o} b_{i}}{\hbar}-i[\frac{V_{cl}}{\hbar}+U_{o}c^{\dagger} c]J_{o}b_{i}-i\frac{U}{\hbar}b^{\dagger}_{i} b_{i} b_{i}-\Gamma b_{i}+\xi.
\end{equation}

Here $\gamma$ is the cavity decay rate, $\Gamma$ is the condensate decay rate and $\xi$ is the classical thermal noise operator of the BEC and has the following correlation function at temperature $T$. $<\xi(t) \xi(t')>$=$<\xi^{\dagger}(t) \xi(t')>$=$<\xi^{\dagger}(t) \xi^{\dagger}(t')>=0$,$<\xi(t) \xi^{\dagger}(t')>=2 \Gamma (1+2 n_{T}) \delta(t-t')$ and $n_{T}=\coth\left ( \hbar \nu /2 k_{B} T\right )$. The noise operators for the input field obey the following correlation functions: $<c^{in \dagger}(t') c^{in}(t)>=n_{p} \delta(t'-t)$, $<c^{in}(t') c^{in \dagger}(t)>=(n_{p}+1) \delta(t'-t)$.The thermal noise input for the BEC is provided by the thermal cloud of atoms.

We now write each canonical operator of the system as a sum of its steady state mean value and a small fluctuation with zero mean value i.e $c\rightarrow c_{s}+c$ and $b_{i} \rightarrow (\sqrt{N/M}+b)$ and linearize to obtain the following Heisenberg-Langevin equations for the fluctuation operators.

\begin{equation}
\dot c=-i \Delta c-i U_{o} J_{o} \sqrt{N} c_{s} (b+b^{\dagger})+\sqrt{2 \gamma} c^{in}-\gamma c
\end{equation}

\begin{equation}
\dot b= -i [\nu+2 U_{eff}]b -i U_{eff} b^{\dagger}-i g_{c}(c^{\dagger}c_{s}+c_{s}^{\dagger}c)-\Gamma b+ \xi,
\end{equation}

Here, $U_{eff}=\dfrac{U N}{M \hbar}$, $g_{c}=U_{0}J_{0}\sqrt{N}|{c}_{s}|$ , $\nu= U_{0}J_{0}|{c}_{s}|^{2}+\dfrac{V_{cl}J_{0}}{\hbar}+\dfrac{E_{0}}{\hbar}$, $\Delta=\Delta_{c}+U_{0}NJ_{0}$ is the detuning with respect to the renormalized resonance.  $N$ is the total number of atoms in $M$ sites. As before, we assume negligible tunneling ($J=E=0$) and hence we drop the site index $j$ from the atomic operators. We now make the transformations $c=\tilde c e^{i \theta}$, $c_{s}=\tilde{c}_{s} e^{-i \theta}$ and $b=\tilde{b} e^{i \theta}$. Neglecting the fast rotating terms, we get the following equations

\begin{equation}\label{fluc_1}
\dot{\tilde c}=-i \Delta c-i g_{c} \tilde{c}_{s} \tilde{b} e^{i \theta}+\sqrt{2 \gamma} \tilde{c}^{in}-\gamma \tilde{c}
\end{equation}

\begin{equation}\label{fluc_2}
\dot{\tilde b}=-i (\nu+2 U_{eff}) \tilde b-i g_{c} \tilde{c}_{s}^{\dagger} \tilde{c} e^{i \theta}- \Gamma \tilde b+ \tilde \xi.
\end{equation}

If we assume that decay rate $\gamma$ of the cavity field is very large so that we can adiabatically eliminate the dynamics of the cavity mode. Consequently we have the steady state value of $\tilde c$.

\begin{equation}\label{steady_1}
\tilde{c}= \frac{1}{\gamma+i \Delta}\left( -i U_{o} J_{o} \sqrt{N} \tilde{c}_{s} \tilde{b} e^{-i \theta}+\sqrt{2 \gamma \tilde{c}^{in}}\right).
\end{equation}

On substituting \ref{steady_1} in \ref{fluc_2} and assuming $\theta=-\pi/2$ for simplicity we get

\begin{equation}\label{fluc_3}
\dot{\tilde b}=-i (\nu+2 U_{eff}) \tilde b- \frac{\beta}{\gamma+i \Delta} \tilde{b}-\frac{\sqrt{2 \gamma \beta}}{\gamma+i \Delta} \tilde{c}^{in} - \Gamma \tilde b+ \tilde \xi.
\end{equation}

Here $\beta=g_{c}^{2}|c_{s}|^{2}$. If $\omega_{m}=\nu+2 U_{eff}>> \Gamma$, the statistics of the input light field can be transferred to the collective density excitations of the BEC.

\section{Entangling two Bose-Einstein condensates}

Here we now utilize the results obtained in the previous section to show that how we can transfer entanglement of a pair of quantum correlated light fields into a pair of BECs in two independent cavity which are physically separated. The scheme is shown in Figure 2. The quantum correlated light fields are generated by a NOPA. The output fields are frequency degenerate but polarization nondegenerate. The coupling between the two intracavity field modes $c_{1}$ and $c_{2}$ can be represented by $i \hbar \chi (c_{1}c_{2}-c_{1}^{\dagger}c_{2}^{\dagger})$. Here $\chi$ is the coupling strength that is proportional to the nonlinear susceptibility of the intracavity medium and the intensity of the coherent pump field. We take $k_{c}$ as the damping rate of the cavity modes. The equations of motion for the mode operators can be written as

\begin{equation}
\dot{c_{1}}=-k_{c}c_{1}-\chi c_{2}^{\dagger}+\sqrt{2 k_{c}} c_{1}^{in},
\end{equation}

\begin{equation}
\dot{c_{2}}=-k_{c}c_{2}-\chi c_{1}^{\dagger}+\sqrt{2 k_{c}} c_{2}^{in},
\end{equation}

where $c_{1}^{in}$ and $c_{2}^{in}$ are the vacuum input fields of the two cavity modes of NOPA. The output fields
from the NOPA follows from the boundary conditions

\begin{equation}
c_{1}^{out}=\sqrt{2 k_{c}} c_{1}-c_{1}^{in},
\end{equation}

\begin{equation}
c_{2}^{out}=\sqrt{2 k_{c}} c_{2}-c_{2}^{in},
\end{equation}

\begin{figure}[t]
\hspace{-0.0cm}
\includegraphics [scale=0.5]{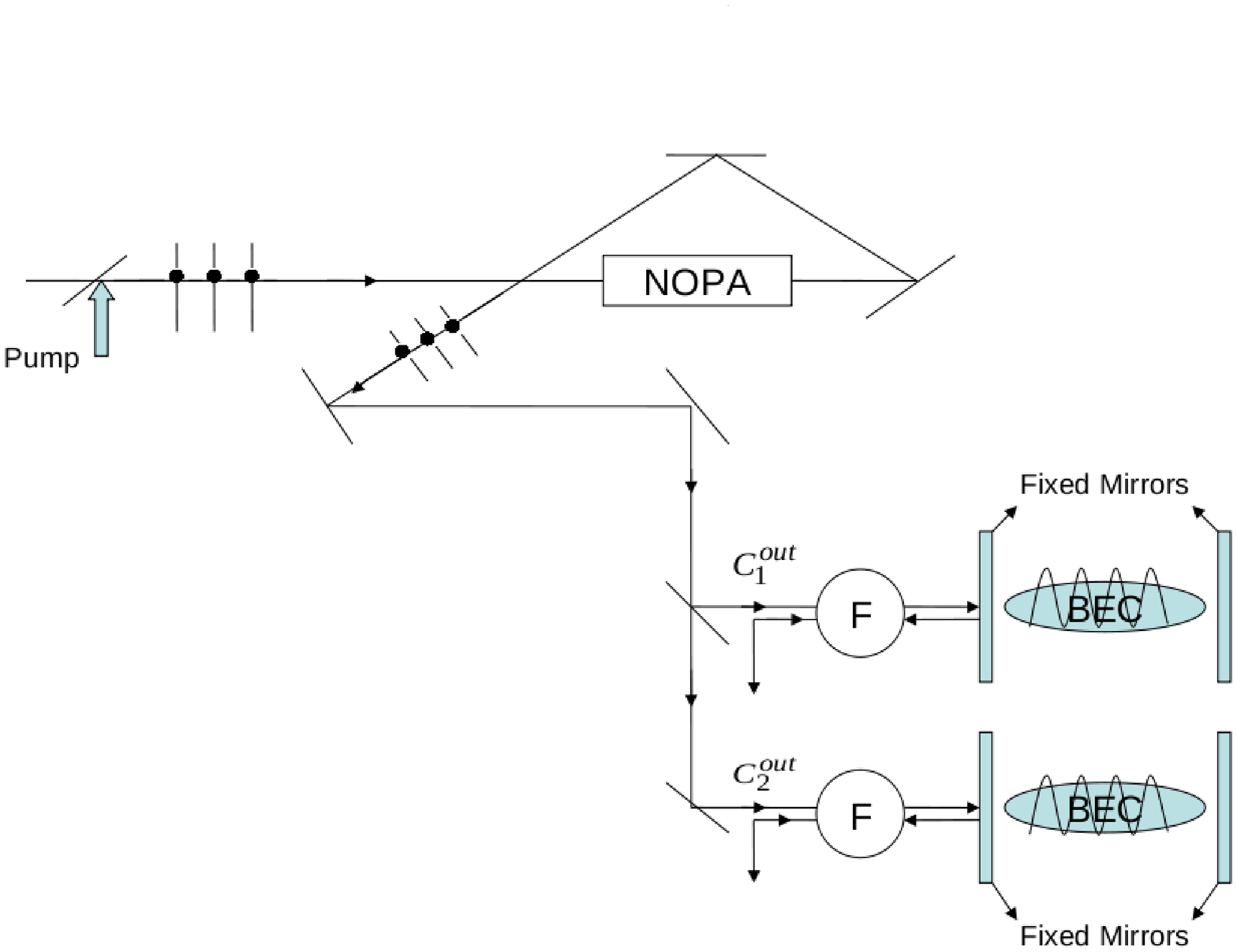}
\caption{Schematic setup to prepare an EPR state of the collective density excitations of two BECs. The output modes from the NOPA are entangled and are incident on the two cavities. The quantum entanglement of the two modes are then transferred to the two BECs.   }
\label{2}
\end{figure}

The quadrature amplitudes $X_{i}^{out}=c_{i}^{out}+c_{i}^{out \dagger}$, $Y_{i}^{out}=i(c_{i}^{out}-c_{i}^{out \dagger})$, $i=1,2$ in Fourier space are found as:

\begin{equation}
X_{1}^{out}(\omega)+X_{2}^{out}(\omega)=\frac{k_{c}-\chi+i \omega}{k_{c}+\chi-i \omega}[X_{1}^{in}(\omega)+X_{2}^{in}(\omega)]
\end{equation}

\begin{equation}
Y_{1}^{out}(\omega)-Y_{2}^{out}(\omega)=\frac{k_{c}-\chi+i \omega}{k_{c}+\chi-i \omega}[Y_{1}^{in}(\omega)-Y_{2}^{in}(\omega)]
\end{equation}

The highly correlated light fields from the NOPA are incident on the two space separated cavities with two identical BECs. We can now write the BEC mode operators of the two condensates as

\begin{equation}\label{fluc_3}
\dot{\tilde b}_{1}=-i (\nu+2 U_{eff}) \tilde{b}_{1}- \frac{\beta}{\gamma+i \Delta} \tilde{b}_{1}-\frac{\sqrt{2 \gamma \beta}}{\gamma+i \Delta} \tilde{c}_{1}^{in}- \Gamma \tilde{b}_{1}+ \tilde \xi.
\end{equation}

\begin{equation}\label{fluc_4}
\dot{\tilde b}_{2}=-i (\nu+2 U_{eff}) \tilde{b}_{2}- \frac{\beta}{\gamma+i \Delta} \tilde{b}_{2}-\frac{\sqrt{2 \gamma \beta}}{\gamma+i \Delta} \tilde{c}_{2}^{in} -\Gamma \tilde{b}_{2}+ \tilde \xi.
\end{equation}

Here we assume that the coupling between the NOPA and the cavities is unidirectional. If system parameters are chosen such that $c_{1}^{out}$ and $c_{2}^{out}$ can be regarded as quantum white noise operators, then the variances of the positions ($\Delta X^{2}=<\delta^{2} (\frac{x_{1}+x_{2}}{2})>$) and momenta ($\Delta Y^{2}=<\delta^{2} (\frac{y_{1}-y_{2}}{2})>$) of the two BECs are given by

\begin{equation}\label{variance_1}
\Delta X^{2}= \Delta Y^{2}=\frac{ \beta \gamma}{C} \left( \frac{A^2}{B^2}+1\right)\left\{ \frac{4 \left( \frac{A^2}{B}+B\right)\chi k_{c}- \left( \frac{A^2}{B}+B\right)^{2} (k_{c}+\gamma)+(k_{c}-\chi)^{2}(k_{c}+\chi) }{\left( \frac{A^2}{B}+B\right)(k_{c}+\chi )[(k_{c}+\chi)^{2}-\left( \frac{A^2}{B}+B\right)^{2}]} \right\}+\frac{ \Gamma (1+2 n_{T})}{B}
\end{equation}

\begin{figure}[t]
\hspace{-0.0cm}
\includegraphics [scale=0.8]{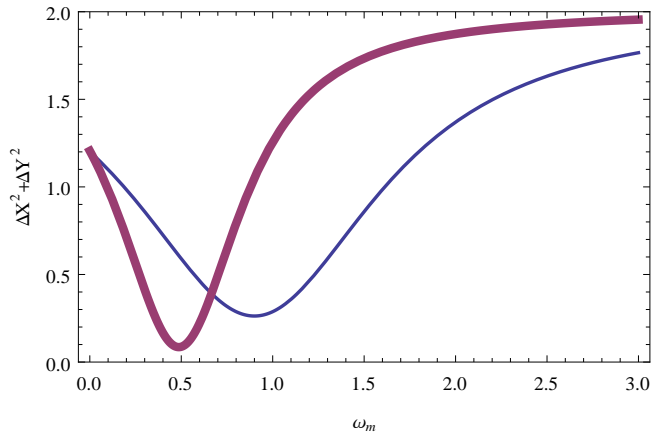}
\caption{A plot of EPR variance $\Delta X^{2}+\Delta Y^{2}$ as a function of $\omega_{m}/\gamma$ for two values of the detuning $\Delta/\gamma=0.3$ (thin line) and $\Delta/\gamma=0.6$ (thick line). The other parameters are $\Gamma/\gamma=0.0001$, $\beta/\gamma=3$, $k_{c}/\gamma=1$, $\chi/\gamma=1$ and $n_{T}=10$. Quantum entanglement is said to exist when $\Delta X^{2}+\Delta Y^{2}<1$. }
\label{3}
\end{figure}

Here, $A=\nu+2 U_{eff}-\Delta \beta/(\gamma^{2}+\Delta^{2})$, $B=\Gamma+\beta \gamma/(\gamma^{2}+\Delta^{2})$, $C=\gamma^{2}+\Delta^{2}$ and $n_{T}=\coth{h \nu/2 k_{B} T}$. The two identical BECs are entangled if $\Delta X^{2}+\Delta Y^{2}<1$. For certain set of parameters, the variances $\Delta X^{2}$ $=$ $\Delta Y^{2}$ $\rightarrow$ $0$ i.e an EPR state in the position and momenta of the two BECs. As evident from Eqn. \ref{variance_1}, the thermal contribution destroys the entanglement. A plot of EPR variance $\Delta X^{2}+\Delta Y^{2}$ as a function of $\omega_{m}/\gamma$ for two values of the detuning $\Delta/\gamma=0.3$ (thin line) and $\Delta/\gamma=0.6$ (thick line) is shown in Fig. 3. Quantum entanglement is said exist when $\Delta X^{2}+\Delta Y^{2}<1$. Clearly we see that significant entanglement is achieved when $A=\nu+2 U_{eff}-\Delta \beta/(\gamma^{2}+\Delta^{2})=0$. Further we observe that for a large detuning $\Delta$, the entanglement is more but the range of $\omega_{m}$ where this entanglement can be achieved is reduced. On the other hand for a smaller detuning, the entanglement is reduced but the range over which entanglement is achieved is enhanced. An EPR state can thus be reached for the collective density excitations of the two BECs by keeping $k_{c}=\chi$ and $A=0$.

To demonstrate that the dynamics investigated here are within experimental reach, we discuss the experimental parameters from \cite{murch,brennecke}: A BEC of typically $10^{5}$ $^{87}Rb$ atoms is coupled to the light field of an optical ultra high-finesse Fabry-Perot cavity. The atom-field coupling $g_{0}=2 \pi \times 10.9 Mhz$ \cite{brennecke} ( $2 \pi \times 14.4 $ \cite{murch}) is greater than the decay rate of the intracavity field $\kappa=2 \pi \times 1.3 Mhz$ \cite{brennecke} ($2 \pi \times 0.66 Mhz$ \cite{murch}).  Typically atom-pump detuning is $2 \pi \times 32 Ghz$. The rate $\Gamma_{b}$ at which atoms are coupled out of the BEC is about $2 \pi \times 7.5 \times 10^{-3} Hz$ \cite{brennecke}. The kinetic energy and potential energy contribution $\nu$ is about $35 kHz$ \cite{brennecke}($49 kHz$ \cite{murch}).The energy of the cavity mode decreases due to the photon loss through the cavity mirrors, which leads to a reduced atom-field coupling. Photon loss can be minimized by using high-Q cavities. Our proposed scheme relies crucially on the fact that coherent dynamics dominate over the losses. It is important that the characteristic time-scales of coherent dynamics are significantly faster than those associated with losses.

\section{Conclusions}

In conclusion, we have proposed a novel scheme to transfer the quantum entanglement between two light fields emerging from a nondegenerate parametric amplifier to the collective density excitations of two Bose Einstein condensates which are physically separated in two independent optical cavities. An EPR state of the collective density excitations can be created by a judicious choice of the parameters. This study could be of use in studying entanglement in macroscopic quantum objects and for high precision metrology.

\end{document}